\documentclass[aps,prl,twocolumn,superscriptaddress,showpacs,amsmath,amssymb]{revtex4-1}
\usepackage{graphicx,bbm}
\usepackage{mathtools}
\usepackage{color}
\newcommand{\be}{\begin{equation}}
\newcommand{\ee}{\end{equation}}
\newcommand{\bea}{\begin{eqnarray}}
\newcommand{\eea}{\end{eqnarray}}

\newcommand{\ket}[1]{{\vert #1 \rangle}}

\newcommand{\ave}[1]{{\langle #1\rangle}}
\newcommand{\ii}{ {\rm i} }
\newcommand{\dd}{ {\rm d} }

\newcommand{\ZZ}{\mathbb{Z}}
\newcommand{\RR}{\mathbb{R}}

\newcommand{\z}{{\rm z}}
\newcommand{\LL}{{\hat {\cal L}}}

\def\vmbb#1{\mathbb{#1}}

\def\tr{{\,{\rm tr}}}
\def\Tr{{\,{\rm Tr}}}

\def\one{\mathbbm{1}}

\def\be{\begin{equation}}
\def\ee{\end{equation}}

\def\TT{\mathbb{T}}
\def\LL{\mathbb{L}}

\def\tr{\,{\rm tr}\,}
\def\ket#1{|#1\rangle}

\def\ave#1{\langle #1 \rangle}
\def\ii{{\rm i}}
\def\z{{\rm z}}

\def\tit#1{}
\newcommand{\half}{{\textstyle\frac{1}{2}}}

\usepackage{color}

\begin{document}
	\title{Lower Bounding Diffusion Constant by the Curvature of Drude Weight}
	\author{Marko Medenjak}
	\affiliation{Faculty of Mathematics and Physics, University of Ljubljana, Jadranska 19, SI-1000 Ljubljana, Slovenia }
	\author{Christoph Karrasch}
	\affiliation{Dahlem Center for Complex Quantum Systems and Fachbereich Physik, Freie Universit ̈at Berlin, 14195 Berlin, Germany }
	\author{Toma\v z Prosen}
	\affiliation{Faculty of Mathematics and Physics, University of Ljubljana, Jadranska 19, SI-1000 Ljubljana, Slovenia }
	\begin{abstract}
		We establish a general connection between ballistic and diffusive transport in systems where the ballistic contribution in canonical ensemble vanishes.  
		A lower bound on the Green-Kubo diffusion constant is derived in terms of the curvature of the ideal transport coefficient, the Drude weight, with respect to the filling parameter. As an application, we explicitly determine the lower bound on the high temperature diffusion constant in the anisotropic spin $1/2$ Heisenberg chain for anisotropy parameters $ \Delta\geq 1 $, thus settling the question whether the transport is sub-diffusive or not. Additionally, the lower bound is shown to saturate the diffusion constant for a certain classical integrable model.
	\end{abstract}
	
	\maketitle

	\date
	\maketitle
	
	{\em Introduction.--} Transport is one of the primary interests in the study of interacting quantum systems. It is still not fully known under which conditions phenomenological laws, such as the current being proportional to the gradient of the charge, apply. Moreover, these diffusion laws are violated by some of the most relevant quantum models, in particular in one dimension (1D) where ideal (ballistic) transport can occur even at finite temperature \cite{giamarchi,FHM03,SPA11}. Experiments in real quasi-1D materials indeed report anomalously high conductivities \cite{hess2001}. In addition to the ballistic Drude peak, the conductivity of these models usually also contains normal, diffusive contributions
\cite{PhysRevB86125118,PhysRevB88205135}.
	
	The wide spectra of transport phenomena exhibited by one dimensional quantum systems can be exemplified by the paradigmatic anisotropic Heisenberg $ XXZ $ model
	\begin{equation}
	H=\frac{1}{4}\sum_{l}\left(\sigma_l^x \sigma_{l+1}^x+\sigma_l^y \sigma_{l+1}^y+\Delta \sigma_l^z \sigma_{l+1}^z+h \sigma_l^z\right).
	\label{eq:XXZ}
	\end{equation}
	Numerical simulations of the spin transport indicate that in the absence of a magnetic field $h$ there are three regimes with distinct transport properties \cite{ZnidaricPRL11}. In particular, numerics suggest that for $|\Delta|<1 $ the spin transport is ideal and that for $|\Delta|=1 $ the system exhibits anomalous behavior. In the regime of $ \Delta>1 $, most of the studies indicate normal (diffusive) transport \cite{ZnidaricPRL11,karraschdif,Steinigewegdif} while others seem to imply insulating behavior \cite{mbp}.
	Adding a nonzero magnetic field $h$ along the $ z $ direction renders the transport manifestly ballistic (ideal) in all regimes \cite{Zotos97,PhysRevB.84.155125}. The ballistic transport coefficient, the Drude weight, is connected to the rate at which the conductivity diverges \cite{IP13} and can be related to the local integrals of motion \cite{Zotos97,Mazur69}, which can be used to stringently prove that transport is ideal for $ |\Delta|<1 $ \cite{ProsenPRL106,PI13,ProsenNPB14,Pereira14}. An ideal transport is in fact the only type understood in the framework of a quasi-particle picture \cite{bertini,PhysRevX.6.041065,bernard_doyon,De_Luca,eiiq}, and the mechanism behind the diffusion in integrable systems is controversial. Diffusion may occur when quasilocal conserved quantities with appropriate symmetry properties are absent, which seems to be the case for 
	$|\Delta|\geq 1 $, $h=0$ in the Heisenberg model. This point of view is supported by a Bethe ansatz calculation and implies the absence of ideal transport at the isotropic point \cite{Carmelo}. A rigorous lower bound on the infinite temperature diffusion constant from the existence of quadratically extensive (nonlocal) almost conserved operators should also be noted \cite{ProsenPRE14}.
	
	In this Letter we consider situation where the Drude weight vanishes due to existence of a $\ZZ_2$ symmetry, such as parity-hole or spin reversal.
	We obtain a lower bound on the diffusion constant which is proportional to the curvature of the Drude weight with respect to a symmetry-breaking parameter (filling fraction, or magnetization) and thus establish a connection between these two transport coefficients.  The main step is the observation that, even though in the thermodynamically dominant ($\ZZ_2$-symmetric) `half-filled' subspace there is no ballistic transport, the ballistically spreading excitations from 
	particle-number (or magnetization) subspaces that have thermodynamically vanishing relative weight generically yield a finite contribution to the diffusion constant. 
	
	Besides providing the bound which is applicable to numerous integrable models \cite{Faddeev_arxiv,KorepinBook}, we identify the mechanism by which an integrable model can exhibit normal transport. Moreover, the fundamental relation between the two transport coefficients is useful since the calculation of the diffusion constant is typically intractable, but we have a good handle on how to compute the curvature of Drude weight, either in terms of local charges \cite{PI13} or by employing the generalized hydrodynamics \cite{eiiq}. Using the bound we show that the finite-temperature spin transport in regime $|\Delta|\geq1 $ of $XXZ$ model is not sub-diffusive, thus settling the issue  outlined by \cite{mbp}. This  is achieved by employing local and quasilocal conserved quantities \cite{IMP15,IMPZ}, and providing a closed form expression for the lower bound on the Drude weight curvature, including all local integrals of motion. To determine whether the quasilocal charges give a correct value for the curvature of the Drude weight, the results are compared to tDMRG simulations \cite{Schollwck201196,PhysRevLett108227206}. Interestingly enough, in the isotropic case $\Delta=1$ the agreement is not perfect. While this can most likely be attributed to the finite accessible timescales of tDMRG simulations, it leaves open the possibility of non-analytical behavior of Drude weight at half-filling, which could imply anomalous spin transport. In addition, the lower bound saturates the diffusion constant for a certain classical integrable model \cite{MKP}.
	
	{\em Transport and parity.--}
	In this subsection we precisely define our setting and discuss the relevance of symmetries with regards to the transport properties.
	We consider dynamics induced by a periodic local Hamiltonian $ H_n $ on the chain of length $ 2 n $. We assume that $H_n$ and any quasilocal conserved charge $ Q_n $ commutes with an extensive ultra-local particle number (or magnerization) operator $ M_n $,
	$ \left[H_n,M_n\right]=0$. Additionally, we require $ H $ to be space reflection invariant.	By an extensive local operator $ A_n $ we mean a sum
	$
	A_n=\sum_{l=-n+1}^{n}a_l,
	$
 	where $a$ is a local operator density supported on sites $[0,|a|-1]$, $a_l$ denotes a shift of the density by $l$ sites to the right and $2n$ periodicity is taken as $n \equiv -n $. Similarly, quasilocality  denotes an extensive operator comprised of local densities which can have infinite support, however an appropriate norm \cite{IMPZ} of the terms should decrease exponentially with their support.
	We further assume that Hamiltonian together with any local or quasilocal conserved quantity $ Q_n $ is symmetric with respect to a $\ZZ_2$ parity (particle-hole) transformation $ S_n $
	\begin{equation}
	\left[Q_n,S_n\right]=0.
	\label{sym}
	\end{equation}
	For instance, in $ XXZ $ model (\ref{eq:XXZ}) the parity  corresponds to the spin-flip operator $ S_n=(\sigma^x)^{\otimes 2 n} $ and $M_n =\sum_{l=-n+1}^n \frac{1}{2}(\sigma^z_l+1)$. Note that the spin or particle current $ J_n $, associated with the magnetization operator $ M_n $ by the continuity equation, is odd under parity $\{J_n,S_n\}=0$, which can be easily understood since, if the spins are flipped the spin current will flow in the opposite  direction. 
	
	To precisely define the diffusion constant we first introduce the Kubo-Mori inner product 
	\begin{equation}
	\langle A,B \rangle^{\beta}_n= \frac{1}{\beta}\int_{0}^{\beta} \dd \lambda\ave{ A^{\dagger} e^{-\lambda H_n}B e^{\lambda H_n}}^\beta_n.
	\end{equation} 
	where $\ave{A}^n_\beta := \tr(e^{-\beta H_n} A)/\tr(e^{-\beta H_n })$ is the canonical thermal expectation value. Similarly we define the projected version of the inner product
	$\ave{A,B}^{\beta,x}_n$, on the subspace of particle number conserving operators, by projecting a canonical expectation value to
	a fixed filling (magnetization) sector $ x $
	\be
	\ave{A}^{\beta,x}_n = \frac{\ave{A P^{(x+1)n}_n}^\beta_n}{\ave{P^{(x+1)n}_n}^\beta_n}.\ee
	Here $ P^m_n $ is a projection operator to an eigenspace of  $ M_n $  with eigenvalue $ m $: $M_n = \sum_m m P^m_n$, $P^{m}_n P^{m'}_n = P^m_n \delta_{m,m'}$. For instance, $ x=0 $ denotes half-filling and $ x=\pm 1 $ corresponds to maximally polarized states.
	
	Kubo linear response formula for the real part of d.c. conductivity is related to the diffusion constant through the Einstein relation $ \sigma(\beta)=\chi(\beta)  \mathcal{D}(\beta)  $, where $\mathcal{D}(\beta) = \lim_{T\to\infty}\lim_{n\to\infty}\tilde{\mathcal{D}}(\beta)$ with
	\begin{equation}
	\tilde{\mathcal{D}}(\beta) =  \frac{\beta}{4 n \chi(\beta)}
	\int_{-T}^{T} \dd t \, \langle \tau^n_t(J_n), J_n\rangle_n^\beta  .
	\label{eqn:Kubo}
	\end{equation}
	Here $J_n=\sum_{l=-n+1}^{n}j_l$ is an extensive current operator with a local density $ j $, and $ \tau^n_t(j)=e^{\ii H_n t} j e^{-\ii H_n t} $ its dynamics. For instance in the Heisenberg model the local current reads $ j=-\frac{i}{2}( \sigma^+_0 \sigma^-_1 -  \sigma^-_0 \sigma^+_1) $
	The static susceptibility is $
	\frac{\chi(\beta)}{\beta}=\lim_{n\to\infty}\frac{\langle M_n^2 \rangle^\beta_n-(\langle M_n\rangle^{\beta}_n)^2}{2n}
	$ and $ \tilde{\chi}=\lim_{\beta\to 0} \frac{\chi(\beta)}{\beta} $ its infinite temperature limit.
	For the Heisenberg $ XXZ $ model we have $\tilde{\chi}=\frac{1}{4} $.
	
	Note that the existence of parity antisymmetric quasilocal conserved operators $ Z_n $ \cite{ProsenPRL106} implies the divergence of conductivity due to the finiteness of Drude weight \cite{Zotos97,Mazur69}, 
	defined as $ D=\lim_{T\to\infty}\lim_{n\to\infty} \tilde{D} $ with
	\begin{equation}
	\tilde{D}(\beta,x)=
	\frac{\beta}{4 T n}\int_{-T}^{T} \dd t  \langle \tau^n_t(J_n), J_n\rangle_n^{\beta,x}.
	\label{eqn:Drude}
	\end{equation}
	Drude weight $ D $ can be bound from below by conserved operator $ Z_n $
	\begin{equation}
	D(\beta,x)\geq \frac{\beta}{2} \lim_{n\to\infty}  \frac{|\langle J_n, Z_n \rangle_n^{\beta,x} |^2}{2 n\ \langle Z_n, Z_n \rangle_n^{\beta,x}}.
	\label{lbdrud}
	\end{equation}
	Provided that there are no symmetry restrictions and that $ Z_n $ is extensive the above bound is expected to be nonvanishing. In contrast, if all local integrals of motion transform as assumed in \eqref{sym} and the ensemble is parity symmetric, i.e. $ x=0 $, the bound \eqref{lbdrud} is zero. This gives rise to the possibility of diffusion in integrable systems. If, however, $ x\neq 0 $ the Drude weight \eqref{eqn:Drude} is expected to be finite \cite{Zotos97,PhysRevB.84.155125}. 
	\smallskip
	\noindent
	
	{\em Lower bound on diffusion constant.--}
	In what follows the contributions to diffusion constant \eqref{eqn:Kubo} from the ballistic sectors, $ x\neq 0 $, are shown to be finite.
	Our main result is the relation between the diffusion constant $ \mathcal{D} $ and the curvature of the Drude weight
	\begin{equation}
	\mathcal{D}(\beta)\geq \frac{1}{8\beta v_{\rm LR}\chi(\beta) f_1(\beta)} \frac{\partial^2}{\partial x^2}D(\beta,x)\Bigr|_{x=0},
	\label{diflb}
	\end{equation}
	where $ v_{\rm LR} $ is a Lieb-Robinson velocity \cite{Lieb1972, bratteli2012operator}, and $ f_1(\beta)=\lim_{n\to\infty}\frac{1}{4n} \frac{\partial^2}{\partial x^2} F_n(\beta,x)|_{x=0}  $ is a second derivative of free energy density 
	at half-filling.
	
	The connection between the finite time and finite system-size diffusion constant $ \tilde{\mathcal{D}} $ \eqref{eqn:Kubo} and the Drude weight $ \tilde{D} $ \eqref{eqn:Drude} is apparent
	\begin{equation}
	\label{dif3}
	\tilde{\mathcal{D}}(\beta)=\frac{T}{\chi(\beta)}\sum_x \langle P_n^{(x+1)n} \rangle ^\beta_n\tilde{D}(\beta,x).
	\end{equation}
	To obtain the expression \eqref{dif3}, we inserted the resolution of the identity $ \one=\sum_x P_n^{(x+1)n} $ into the expression \eqref{eqn:Kubo}, renormalizing each term by $\langle P_n^{(x+1)n} \rangle$.
	The central observation is that the diffusion constant can be defined as a single scaled limit $ T\to\infty $, by replacing the size of the system $ 2n $ with $ 2v T $, where the velocity $ v $ should be greater than the Lieb-Robinson velocity $ v_{\rm LR} $ \cite{Lieb1972, bratteli2012operator}, i.e. the maximal velocity with which the information can travel through the spin chain.
	
	\smallskip
	
	\begin{figure*}
		
		\includegraphics[width=0.95\linewidth]{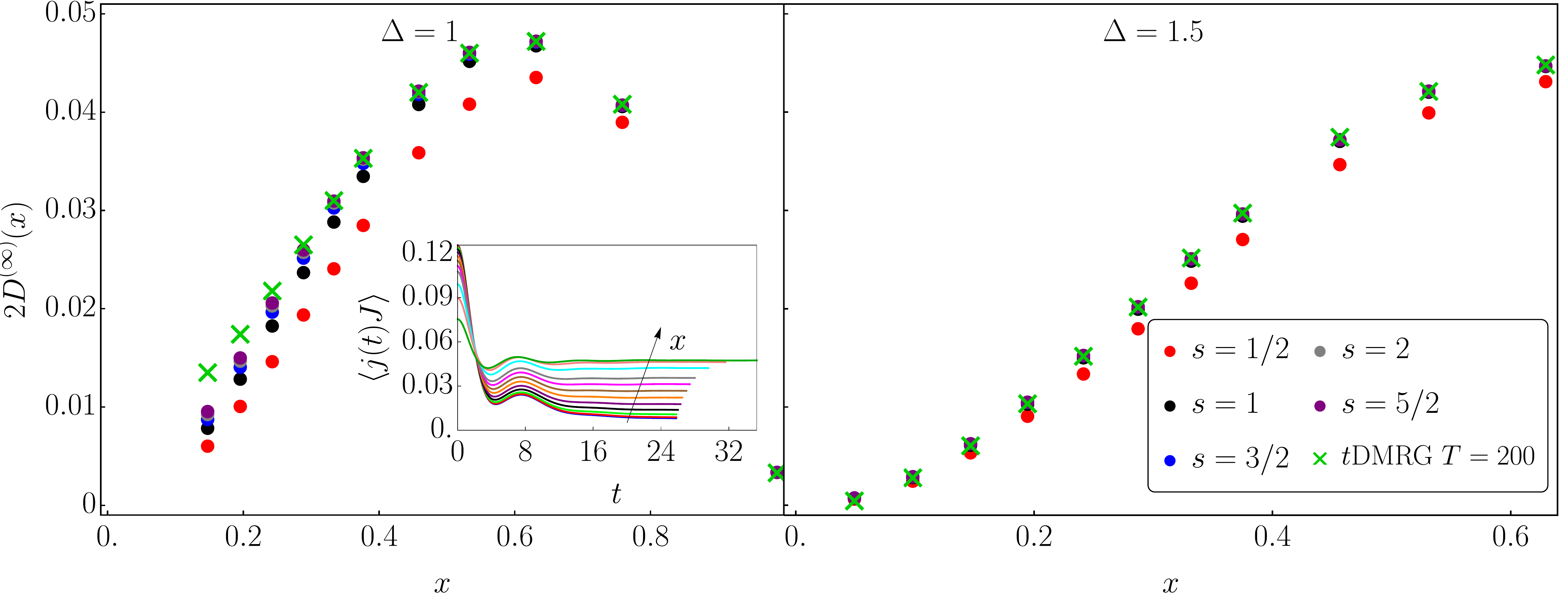}
		\center
		\caption{
			\label{fig1} 
			Optimized high-temperature Drude weight lower bound $g(0,x)$ obtained from a finite number of quasilocal charges for $ s=\half $ up to $ s=\frac{5}{2} $ is plotted for $\Delta=1$ (a) and $\Delta=1.5$ (b). Results are compared to the finite time tDMRG result of current-current autocorrelation function at temperature $ T=200 $ (green crosses). Inset in the left figure shows real time tDMRG data for $ \Delta=1 $ with filling $ x $ increasing from bottom to top, starting from half-filling.}
		
	\end{figure*}
	
	\noindent
	This is a consequence of the Lieb-Robinson theorem \cite{Lieb1972,bratteli2012operator,Verstraete} and the clustering property of spatio-temporal autocorrelation function \cite{IP13} (see Sec.~A of \cite{sup}).
	Setting $ n\equiv v T $, and expanding the Drude weight for large times $ \tilde{D}=D(\beta,x)+\frac{1}{T}D_1(\beta,x)+{\cal O}(1/T^2) $, the scaling contribution  $ D_1(\beta,x) $ and the ballistic  contribution $ D(\beta,x) $ can be identified. Note that $D_1(\beta,x)$ in fact takes a form of a Green-Kubo expression
	for the diffusion constant in the presence of convective term \cite{spohn}, namely by the current operator $J(t)$ replaced by $\tilde{J}(t) = J(t) - \frac{1}{2T}\int_{-T}^T {\rm d}t\, J(t)$, before taking $T\to\infty$, and can be shown to be manifestly nonnegative.
	In what follows we take into account only the ballistic contribution. For infinite temperature, the statistical weights can be calculated explicitly $\langle P^m_{vT} \rangle^0_{vT}=\frac{1}{2^{2vT}}\binom{2vT}{m}$. Expanding the Drude weight in $x$ around half-filling $x=0$ and taking into account only the leading contribution $ D(0,x)\sim\half \frac{\partial^2}{\partial x^2} D(0,0) x^2 $, the result
	\begin{equation}
		\label{lb}
		\mathcal{D}(0)\geq \frac{1}{4\tilde{\chi}v} \frac{\partial^2}{\partial x^2}D^{(\infty)}(x)\big\vert_{x=0},
	\end{equation}
	is obtained, with $ D^{(\infty)}(x)=\lim_{\beta\to0}\frac{D(\beta,x)}{\beta} $. Higher order contributions can be shown to vanish (see Sec.~B of \cite{sup}).
	
	Obtaining the finite temperature bound is straight forward, after making a few assumptions. First of all introducing the filling-dependent free energy function
	\begin{equation}
	\beta F_n(x,\beta)=-\log \tr (P_n^{(x+1)n} e^{-\beta H_n}),
	\end{equation}
	disregarding the contributions to the free energy function from the states that are sufficiently far away from half-filling (see sec. C of \cite{sup}), the statistical weights of sectors can be calculated as $\langle P^{(x+1)n}_n\rangle_n^\beta \propto e^{-f_1(\beta)x^2 2 n}$ .
	Lastly, a summation over filling sectors in expression \eqref{dif3} can be replaced by integration yielding the main result \eqref{diflb}  (for details see Sec.~C of \cite{sup}).
		
	The dependence of the lower bound on the velocity $ v $ might seem puzzling at first, since the diffusion constant $ \mathcal{D} $ is independent of $ v $, provided that  $ v \geq v_{\rm LR} $. However, one can quickly see that the scaling contribution $\sum_{x} P^{(x+1)v T}_{v T} D_1(\beta,x) $ to expression (\ref{dif3}), which has been disregarded in the lower bound, depends on the velocity $v$ as well. Note that in the limit $v\to\infty$ this latter expression contains the
	entire diffusion constant, so our lower bound vanishes. We thus expect that the optimal bound, without further considerations, is achieved for $v=v_{\rm LR}$.
	 
	{\em Example: Heisenberg model.--}
	Here we obtain a bound on diffusion constant in $ XXZ $ model, by employing the Mazur inequality to lower bound the curvature of the Drude weight.
	We begin by noting that a set of quasilocal charges is generated by logarithmic derivatives of transfer matrices $ T_s(\lambda) $ \cite{IMP15,IMPZ}
	\begin{equation}
	X_s(\lambda)=\partial_{\lambda} \log T^+_s(\lambda),
	\label{charges}
	\end{equation}	
	 where $ \lambda $ is a spectral parameter, representation (spin) parameter $ s $ takes half integer values, parameter shift is denoted by $ f^+(\lambda)=f(\lambda+\ii \frac{\gamma}{2}) $, and $ \gamma> 0 $ parametrizes the anisotropy as $ \Delta=\cosh(\gamma) $.	For simplicity we consider here only the high temperature limit $\beta=0$.
	 To obtain an optimal bound in the filling sector $ x $ we introduce functions $ h^x_s (\lambda) $ expressing the quasilocal charge $ Q^x=\sum_s \int \dd \lambda\ h^x_s(\lambda) X_s(\lambda) $ and study the continuous version of the least-square problem. An optimal function $ h^x_s(\lambda) $ can be obtained by minimizing the expectation value of the square of the operator
	 \begin{equation}
	 B^x= \lim_{T\to\infty}\frac{1}{T}\int_0^T \!\!dt\,\tau_t(J) -\sum_s \int h^x_s(\lambda)X_s(\lambda) d\lambda
	 \end{equation}
	 in magnetization sector $ x $. This yields a set of coupled integral equations for the functions $ h^x_s(\lambda) $
	\begin{equation}
	\sum_{s=1/2}^\infty \int d \lambda\ h^x_s(\lambda)K^x_{s,s'}(\lambda,\mu)=J^x_{s'}(\mu)
	\label{sys}
	\end{equation}
	where the expressions for the kernels and overlaps read
	\begin{eqnarray}
	K^x_{s,s'}:=K^x_{s,s'}(\lambda,\mu)&=&\lim_{n\to\infty}\frac{1}{n}\bigl(\langle X_s(\lambda) X_{s'}(\mu )\rangle_n^{0,x}-\nonumber
	\\	
	&-&\langle X_s(\lambda) \rangle_n^{0,x}\langle X_{s'}(\mu ) \rangle_n^{0,x}\bigr),
	\label{ker}
	\\
	J^x_s(\lambda)&=&\lim_{n\to\infty} \langle j X_s(\lambda)\rangle_n^{0,x}.
	\label{ov}
	\end{eqnarray}
	The charge $ Q^x$ can be plugged into the Mazur inequality, yielding
	\begin{equation}
	D(0,x) \geq \sum_{s,s'=\half}^\infty\int d\lambda \int d\mu\  K^x_{s,s'} h^x_s(\lambda) \bar{h}^x_{s'}(\mu).
	\end{equation}
 	In practice, the kernels \eqref{ker} and overlaps \eqref{ov} are calculated in grand-canonical ensemble with chemical potential $ \kappa $,  related to expectation value of magnetization density as $ x=\tanh \kappa $. 
	To calculate the kernels and overlaps one can employ explicit matrix product representation of the charges (for further details see Sec.~D of \cite{sup}).		
	Taking into account only local charges, the problem of obtaining an optimal lower bound in the vicinity of half-filling can be reduced to the infinite tridiagonal Toeplitz system and solved exactly (for details see section E of \cite{sup}). An optimized bound including all strictly local charges, relying on conjectured expression for the kernel $ K^0_{s,s'} $, reads
	\begin{equation}
	\mathcal{D}(0)\geq \frac{\cosh(\gamma)}{3\ v_{\rm LR}}\left(e^{-\gamma}+\frac{2 \sinh \gamma}{\sqrt{1+e^{2\gamma}+e^{4\gamma}}+2+e^{2\gamma}}\right).
	\label{lrb}
	\end{equation}
	To obtain an optimized bound, including quasilocal charges, we took a finite subset of charges and optimized the bound for this subset in thermodynamic limit. From the numerical data on spatio-temporal autocorrelation functions we can estimate the relevant velocity $ v \approx 1 $ for $ \Delta=1.5 $ and infinite temperature (see sec.~F of \cite{sup}). Using this estimate the lower bound on diffusion constant is a factor of $\sim 3$ smaller than tDMRG result.  
		
	In Fig.~\ref{fig1} we plot the comparison between the lower bounds on Drude weight obtained by including different number of families of quasilocal charges and tDMRG results for Drude weight for $ \Delta=1 $ and $ \Delta=1.5 $. In case of $ \Delta=1.5 $ the lower bound almost perfectly saturates the Drude weight. Using only the strictly local charges ($s=1/2$) already gives very good estimate of the behavior in the vicinity of $ x=0 $. The results for isotropic point are more intriguing.
	While for large enough $ x $ an agreement is once again almost perfect, the discrepancies begin to show in the vicinity of $ x=0 $. This can be attributed either to finite accessible times of tDMRG simulation or to the non-optimal lower bound. While the first explanation seems more likely, since as we approach half-filling the relaxation times seem to increase, the second explanation is more attractive as it could imply non-analytic behavior of Drude weight at half-filling and in turn super-diffusive transport \cite{ZnidaricPRL11}.
	
	{\em Saturated lower bound for a classical model.--}
	One of the primary questions that arises from the present study is when and if the lower bound presented here saturates the diffusion constant or not. Recently, a classical model of hard-core interacting charged particles was proposed, for which the diffusion constant and the Drude weight were calculated exactly \cite{MKP}. Interestingly enough, the lower bound presented here saturates the diffusion constant in case of deterministic scattering of the classical particles, and provides a strict lower bound for the stochastic version of the model.
	
	{\em Summary.--} We investigated the connection between the diffusion constant and the local conservation laws. While ballistic transport could be expected in integrable models with infinite number of conserved charges, this may not occur when considering ensembles with additional $\ZZ_2$ symmetries. These include important examples such as Gibbs ensembles, or even space reflection invariant generalized Gibbs ensembles at half-filling. Ballistic transport in these cases requires special set of charges which only appear for special examples with enhanced symmetries, such as $XXZ$ model with rational $\frac{1}{\pi}\arccos\Delta$ \cite{PI13}. Our results settle the question whether the transport in $ XXZ $ model is diffusive or sub-diffusive, offering an explanation of discrepancies between the results of Refs.~\cite{ZnidaricPRL11,karraschdif,Steinigewegdif} and \cite{mbp} which could be attributed to the latter study specializing to half-filling and thus disregarding the contributions of ideal transport away but near half-filling.  An important open question is  whether the behavior of Drude weight w.r.t. filling is analytic,
	in particular at the isotropic point $\Delta=1$, since the opposite would imply the divergence of diffusion constant and super-diffusive, sub-ballistic transport consistently with the result of \cite{ZnidaricPRL11}. Interestingly enough our bound saturates the diffusion constant of a classical hard-core interacting lattice gas. Further objectives along the way of the paper are to optimize a lower bound by including both quasilocal and quadratically extensive charges and calculating the bound for other models.
	
	\smallskip
 We would like to thank E. Ilievski, F. Heidrich-Meisner, and H. Spohn for useful remarks on the manuscript. MM thanks K. Klobas and L. Zadnik for fruitful discussions.
 The work has been supported by ERC grant OMNES and Slovenian Research Agency grant N1-0025 and programme P1-0044. CK is supported by the DFG via the Emmy-Noether program under KA 3360/2-1.

	\bibliography{dif_lb}

	\newpage
	\begin{widetext}
		
		\pagebreak
		
		\begin{center}
			\textbf{ \large {\em Supplemental material}:\\
			Lower Bounding Diffusion Constant by the Curvature of Drude Weight}
		\end{center}


	\section{A: Modifying the expression for the diffusion constant.}
	First of all we will demonstrate how to obtain the single scaled expression for diffusion constant
	\begin{equation}
	\label{scaled_dif}
	\mathcal{D}(\beta,x)=\lim_{T\to\infty} \frac{\beta}{2 vT \chi(\beta)}
	\int_{-T}^{T} \dd t \, \langle \tau^{v T}_t(J_{vT}), J_{vT}\rangle_{vT}^\beta 
	\end{equation}
	for $ v\geq v_{\rm LR} $ from the initial definition. Taking into account exponential clustering property of spatio-temporal autocorrelation function \cite{IP13} 
	\begin{equation} 
	\langle \tau^n_t(j) j_x\rangle \leq \| j\|^2 \min(1,\exp(-\lambda(|x|-|j|-v_{\rm LR} t))),
	\end{equation} for some $\lambda,v_{\rm LR}>0$, enables us to take into account only $ J_r $ contribution of extensive current in the definition of diffusion constant,  provided that the integration time is bounded by $ T\leq\frac{r}{\, v_{\rm LR}}$. Furthermore in next paragraph we will show that $ \tau_t^n $ can be replaced by $ \tau_t^r $ as a consequence of Lieb-Robinson theorem. We make an additional assumption that Gibbs state $ e^{-\beta H_n} $ in the initial definition of diffusion constant can be replaced by $ e^{-\beta H_r} $, which is justified in the limit of large $r$ due to finiteness of thermal correlation length and that the imaginary time propagation of the current density $ j $  is a quasilocal density. Putting all of the above observations together yields the expression \eqref{scaled_dif}.
	
	Here we demonstrate, that the dynamics of a local observable induced by Hamiltonian corresponding to periodic boundary condition can be replaced by Hamiltonian for open boundary conditions, provided that the evolved observable is localized far enough from the boundary. Let $ H_n^o $ and $ H_n^p $ correspond to open and periodic boundary conditions respectively and $ h_n^b $ denote the boundary terms
	\begin{equation}
	H_n^o=H_n^p-h_n^b.
	\end{equation}
	The time evolution corresponding to periodic Hamiltonian can be expanded in terms of open Hamiltonian and the boundary part using Suzuki-Trotter decomposition
	\begin{equation}
	e^{\ii H_n^p t}j e^{-\ii H_n^p t}=\lim_{k\to\infty}(e^{\frac{\ii t}{k}h_n^b}e^{\frac{\ii t}{k}H_n^o})^k j (e^{-\frac{\ii t}{k}H_n^o} e^{-\frac{\ii t}{k}h_n^b})^k.
	\end{equation}
	Commuting operators $ e^{-\frac{\ii t}{k} h_n^b} $ to the left one obtains
	\begin{equation}
	\| e^{\ii H_n^p t}j e^{-\ii H_n^p t}-e^{\ii H_n^o t}j e^{-\ii H_n^o t}\|=\|\lim_{k\to\infty}\sum_{\tilde{k}=1}^{k} (e^{\frac{\ii t}{k}h_n^b}e^{\frac{\ii t}{k}H_n^o})^{k-\tilde{k}}e^{\frac{\ii t}{k} h_n^b} \left[e^{\frac{\ii\tilde{k} t}{k} H_n^o }j e^{-\frac{\ii\tilde{k} t}{k} H_n^o },e^{-\frac{\ii t}{k} h_n^b}\right](e^{-\frac{\ii t}{k}h_n^b}e^{-\frac{\ii t}{k}H_n^o})^{k-\tilde{k}}\|
	\end{equation}
	Since the operators on the l.h.s. exist so does the limit on the r.h.s. and using the triangle inequality, unitary invariance of the spectral norm and the property $ \|A B\|\leq \|A\| \|B\| $ we arrive at the following bound
	\begin{equation}
	\| e^{\ii H_n^p t}j e^{-\ii H_n^p t}-e^{\ii H_n^o t}j e^{-\ii H_n^o t}\|\leq\lim_{k\to\infty} \sum_{\tilde{k}=1}^{k}\left\|\left[e^{\frac{\ii\tilde{k} t}{k} H_n^o }j e^{-\frac{\ii\tilde{k} t}{k} H_n^o },e^{-\frac{\ii t}{k} h_n^b}-\one\right]\right\|
	\label{raz1}
	\end{equation} 
	The terms in the above sum can be bounded using Lieb-Robinson bound \cite{Verstraete,Lieb1972,bratteli2012operator} 
	\begin{equation}
	\left\|\left[e^{\frac{\ii\tilde{k} t}{k} H_n^o }j e^{-\frac{\ii\tilde{k} t}{k} H_n^o },e^{-\frac{\ii t}{k} h_n^b}-\one\right]\right\|\leq c \min\{|j|,|h^b_n|\}\|j\|\ \|e^{-\frac{\ii t}{k} h_n^b}-\one\| \exp\left(-\frac{n-|h|-|j|-v \frac{\tilde{k}}{k} t}{\zeta}\right)
	\end{equation}
	yielding the following expression
	\begin{equation}
	\| e^{\ii H_n^p t}j e^{-\ii H_n^p t}-e^{\ii H_n^o t}j e^{-\ii H_n^o t}\|\leq \lim_{k\to\infty}\left(2 c \min\{|j|,|h^b_n|\}\|j\|\ \tfrac{t\|h_n^b\|}{2 k} \tfrac{\exp\left(-\tfrac{n-|h|-|j|-v t}{\zeta}\right)-\exp\left(-\tfrac{n-|h|-|j|}{\zeta}\right)}{1-e^{-\tfrac{t v}{\zeta k}}} \right)
	\end{equation}
	which finally reduces to 
	\begin{equation}
	\| e^{\ii H_n^p t}j e^{-\ii H_n^p t}-e^{\ii H_n^o t}j e^{-\ii H_n^o t}\|\leq  c \min\{|j|,|h^b_n|\}\|j\|\ \tfrac{\zeta \|h_n^b\|}{  v} \left(\exp\left(-\tfrac{n-|h|-|j|-v t}{\zeta}\right)-\exp\left(-\tfrac{n-|h|-|j|}{\zeta}\right)\right).
	\end{equation}
	Provided that we are inside of the Lieb-Robinson cone the above difference is exponentially small in $ n $.
	
	Furthermore, decomposing the Hamiltonian $ H_d^o $ corresponding to the system of size $2 d $ into the contributions from the Hamiltonian $ H_n^o $ acting non-trivially only on the subsystem of the size $2 n $ centered around the origin, the boundary contributions $ h_n^b $ and the complementary part $ H_{d/n}^o $
	\begin{equation}
	e^{\ii H_d^o t}j e^{-\ii H_d^o t}=\lim_{k\to\infty}(e^{\frac{\ii t}{k}H_{d/n}^o}e^{\frac{\ii t}{k}h_n^b}e^{\frac{\ii t}{k}H^o_n})^k j (e^{-\frac{\ii t}{k}H^o_n} e^{-\frac{\ii t}{k}h_n^b}e^{-\frac{\ii t}{k}H_{d/n}^o})^k,
	\end{equation}
	using similar arguments as above, as well as commutativity of $ e^{\ii t H_{d/n}^o} $ with any operator supported on the subsystem corresponding to $ H_n^o $, one can show, provided that we are inside of the Lieb-Robinson cone, that $\| e^{\ii H_d^o t}j e^{-\ii H_d^o t}-e^{\ii H_n^o t}j e^{-\ii H_n^o t}\|$ is exponentially small in $ n $ for any $ d> n $.

	\section{B: Higher moments of Binomial distribution}
	The aim here is to demonstrate that higher even moments $ k>2 $ of binomial distributions vanish
	\begin{equation}
	\lim_{n\to\infty} n \sum_{m=0}^{2n} \frac{1}{2^{2 n}} \binom{2 n}{m} ( \tfrac{m}{n}-1)^k=0.
	\end{equation}
	Using Hoeffding's inequality 
	\cite{Hoeffding:1963} and $| \tfrac{m}{n}-1|\leq 1$ the tail contributions to the above sum can be bound by
	\begin{equation}
	n \left| \sum_{m=0}^{2n} \frac{1}{2^{2n}} \binom{2 n}{m} \left( \frac{m}{n}-1\right)^k-\sum_{m=\lceil n-2n^{1/2+\varepsilon}\rceil}^{\lfloor n+2 n^{1/2+\varepsilon}\rfloor} \frac{1}{2^{2 n}} \binom{2 n}{m} \left( \frac{m}{n}-1\right)^k\right |\leq 2\ n\ e^{-4 n^{2\varepsilon}}.
	\label{dif}
	\end{equation}
	Taking a maximal value of $ ( \tfrac{m}{n}-1)^k $ in the second sum we arrive at the  upper bound for momenta
	\begin{equation}
	n\left(\sum_{m=\lceil n-2n^{1/2+\varepsilon}\rceil}^{\lfloor n+2 n^{1/2+\varepsilon}\rfloor} \frac{1}{2^{2 n}} \binom{2 n}{m} \left( \frac{m}{n}-1\right)^k \right)\leq 2^k n^{-k/2+k \varepsilon+1}.
	\end{equation}
	We see that, provided that $ k\geq 4 $ and $ 0<\varepsilon<\frac{1}{4} $, the above sum as well as the difference \eqref{dif} vanish in the limit $ n\to\infty $.
	\section{C: Replacing a sum with an integral}
	In this subsection we show that the sum over different particle number sectors can be replaced by integration in the thermodynamic limit. The initial expression reads
	\begin{equation}
	\mathcal{D}(\beta)\geq \frac{1}{ \chi(\beta) v_{\rm LR}} \lim_{n\to\infty} n \sum_{m=0}^{2 n} \langle P^m_n \rangle^\beta_n D(\beta,\tfrac{m}{ n}-1).
	\end{equation}
	Note that $ n $ corresponds to $ vT $. Expressing $\langle P^m_n \rangle^\beta_n$ in terms of free energy reads
	\begin{equation}
	\label{lb21}
	\mathcal{D}(\beta)\geq \frac{1}{ \chi(\beta) v_{\rm LR}} \lim_{n\to\infty} \frac{n \sum_{m=0}^{2 n}   \exp(-\beta(f_0(\beta) 2 n+f_1(\beta) (\frac{m}{n}-1)^2 2 n+f_2(\beta)(\frac{m}{n}-1)^4 2 n+...)) D(\beta, \frac{m}{n}-1)}{\sum_{m=0}^{2 n} \exp(-\beta(f_0(\beta) 2 n+f_1(\beta) (\frac{m}{n}-1)^2 2 n+f_2(\beta)(\frac{m}{n}-1)^4 2 n+...))}.
	\end{equation}
	The relative fraction of states outside of the region
	\begin{eqnarray}
	n- (2 n)^{1/2+\varepsilon}<m< \ n+ (2 n)^{1/2+\varepsilon};\ \ \ \ \ 0<\varepsilon<\frac{1}{4}
	\label{reg}
	\end{eqnarray}
	is bounded from above by $ \mathcal{O}(\exp(-A\ n^{2\varepsilon})) $ and we assume that their contribution to the partition function is negligible. Inside of the region \eqref{reg} the lower bound \eqref{lb21} can be approximated by
	\begin{equation}
	\mathcal{D}(\beta)\geq \frac{1}{ \chi(\beta) v_{\rm LR}} \lim_{n\to\infty} \frac{\frac{n^\frac{3}{2}}{n}\sum_{r=-\lfloor 2 n^{\tfrac{1}{2}+\varepsilon}\rfloor}^{\lfloor 2 n^{\tfrac{1}{2}+\varepsilon} \rfloor}   \exp(-2 \beta f_1(\beta) \tfrac{r^2}{ n}) D(\beta, \tfrac{r}{ n})}{\frac{\sqrt{n}}{n}\sum_{r=-\lfloor 2 n^{\tfrac{1}{2}+\varepsilon}\rfloor}^{\lfloor 2 n^{\tfrac{1}{2}+\varepsilon} \rfloor}  \exp(-2 \beta f_1(\beta) \tfrac{r^2}{ n})} \left(1+\mathcal{O}\left(\frac{1}{n^{1-4\varepsilon}}\right) \right),
	\end{equation}
	where we introduced new variable $ r=\ m- n $. The difference between Riemman sums and an integral in denominator can be bounded with
	\begin{equation}
	\left|\frac{\sqrt{n}}{n}\sum_{r=-\lfloor 2 n^{\tfrac{1}{2}+\varepsilon}\rfloor}^{\lfloor 2 n^{\tfrac{1}{2}+\varepsilon} \rfloor}  \exp(-2 \beta f_1(\beta) \tfrac{r^2}{ n})-\int_{-\lfloor 2 n^{ \half+\varepsilon }\rfloor/ n}^{\lfloor 2 n^{ \half+\varepsilon }\rfloor/ n} \dd u \sqrt{n} \exp(-2 n\beta f_1(\beta) u^2)\right|\leq \frac{C}{n^{1/2-\varepsilon}}
	\end{equation}
	and similarly for the enumerator
	\begin{equation}
	\left|\frac{ n^\frac{3}{2}}{n}\sum_{r=-\lfloor 2 n^{\tfrac{1}{2}+\varepsilon}\rfloor}^{\lfloor 2 n^{\tfrac{1}{2}+\varepsilon} \rfloor}  \exp(-2 \beta f_1(\beta) \tfrac{r^2}{ n}) D_n(\beta, \tfrac{r}{ n})- \int_{-\lfloor 2 n^{ \half+\varepsilon }\rfloor/ n}^{\lfloor 2 n^{ \half+\varepsilon }\rfloor/ n} \dd u\ n^\frac{3}{2} \exp(-2 n\beta f_1(\beta) u^2) D(\beta, u)\right|\leq \frac{B}{n^{1/2-\varepsilon}}.
	\end{equation}
	The integration boundaries can be moved to some finite  values since function $ g $ is bounded and the maximum of the function $ \exp(-2 n\beta f_1 u^2) $ on the interval $ u \in [\lfloor 2 n^{ \half+\varepsilon }\rfloor/ n,1] $ is $ A_1 \exp(-A_2\  n^{2\epsilon}) $. This yields the following expression for the lower bound on diffusion constant	 
	 \begin{equation}
	 	\mathcal{D}(\beta)\geq B \lim_{n\to\infty}   \int_{-\eta}^\eta\!\!\!dx\ n^{3/2} e^{-n x^2} D(\beta, \eta x).
		\label{ukor}
	 \end{equation}
	 	where $B= \frac{1}{ \chi(\beta) v_{\rm LR}\sqrt{\pi}} $ and $ \eta=(2 \beta f_1(\beta))^{-1/2} $. In the above calculation we disregarded corrections and calculated the integral in enumerator.
	 	Noticing that $ n^{3/2} e^{-n x^2}=-\frac{1}{2x}\frac{\partial}{\partial_x} \sqrt{n}\ e^{-n x^2} $ and integrating \eqref{ukor} by parts we arrive at the following equality
	 \begin{equation}
	 	\mathcal{D}(\beta)\geq \frac{B}{2} \lim_{n\to\infty} \int_{-\eta}^{\eta}\!\!\!dx\ \sqrt{n}\ e^{-n x^2}\frac{\partial}{\partial x} \frac{D(\beta, \eta x)}{x}.
	 \end{equation}
	 Here the $ n $ dependent part corresponds to regularization of delta distribution, thus yielding the central result from the main text (\ref{diflb}).
	 Calculating $ f_1(0) $ using Stirling approximation we readily recover the infinite temperature result from the main text.

	\section{D: Calculation of norms and overlaps}
	Here we outline a method to calculate norms and overlaps of quasilocal charges and spin current operator in Heiseberg $XXZ$ model for infinite temperature and finite chemical potential, specializing to relevant sector $ \Delta\geq 1 $. Additionally we outline some of the exact and conjectured results.
	
	With a given expectation value of the on-site magnetization $\sigma^z$ we associate a chemical potential $ \kappa $
	\begin{equation}
	x=-\frac{\tr(\sigma^z e^{-\kappa \sigma^z})}{\tr(e^{-\kappa \sigma^z})}=\tanh(\kappa).
	\end{equation}
	The scalar product on the operator space $ {\rm End}(\mathbb{C}^{2^{n}}) \backslash \{ \one\} $ associated with expectation value of magnetization $ x $ reads
	\begin{equation}
	\langle A,B\rangle^x\equiv\frac{\tr(A B^\dagger e^{-\kappa M})}{\tr(e^{-\kappa M})}-\frac{\tr(A  e^{-\kappa M})}{\tr(e^{-\kappa M})}\frac{\tr(B^\dagger e^{-\kappa M})}{\tr(e^{-\kappa M})}.
	\label{scalar_product}
	\end{equation}
	with $ M $ being a global magnetization operator.
	It is advantageous to introduce an orthonormal basis $ {\sigma^z_x, \sigma^+_x,\sigma^-_x} $ w.r.t. inner product \eqref{scalar_product}
	\begin{equation}
	\sigma_x^0=\sigma^0,\
	\sigma_x^\pm=\sigma^\pm \sqrt{\frac{2}{1\pm x}},\
	\sigma_x^z=\frac{\sigma^z +x\ \sigma^0}{\sqrt{1-x^2}},
	\label{ort_basis}
	\end{equation}
	which satisfies additional requirement  
	\begin{equation}
	\tr(\sigma_x ^\alpha e^{-\kappa \sigma^z})=0;\ \  \alpha\in\{+,-,z\} .
	\label{pr2}
	\end{equation}
	In basis \eqref{ort_basis} scalar product \eqref{scalar_product} retains separability of the zero magnetization inner product $ x=0 $ (i.e. operators in the new basis \eqref{ort_basis} with disjoint support are orthogonal).
	Furthermore, any MPO $ A=\sum_{\alpha\in\mathcal{J}} A^\alpha\otimes \sigma^\alpha $ can be  recast as $ A=\sum_{\alpha\in\mathcal{J}} A^{\alpha,x}\otimes \sigma_x^\alpha $ with
	\begin{equation}
	A^{\pm, x}=\sqrt{\frac{1\pm x}{2}}\ A^\pm,\ A^{z,x}=\sqrt{1-x^2}\ A^{z},\ A^{0,x}=A^0-x \ A^z,
	\label{mubasis}
	\end{equation}
	where $ \mathcal{J}=\{0,z,+,-\} $.
	To obtain a lower bound on Drude weight in distinct magnetization sectors, the inner product of the charges
	\begin{equation}
	K^x_{s,s'}(\lambda,\mu)=\lim_{n\to\infty}\frac{1}{n}\langle X_s(\lambda),X_{s'}(\mu )\rangle^{x},
	\label{ker1}
	\end{equation}
	has to be calculated.
         Here $ X_{s}(\lambda) $ are the quasilocal integrals of motion
	\begin{equation}
	X_{s}(\lambda)=-\ii\partial_{\mu}\frac{T^{-}_{s}(\lambda)}{T^{[-2s-1]}_{0}(\lambda)}
	\frac{T^{+}_{s}(\mu)}{T^{[2s+1]}_{0}(\mu)}\Big|_{\mu=\lambda},\qquad \lambda \in \RR.
	\label{eqn:X_product_from}
	\end{equation}
	Transfer matrices are obtained as a partial tensor product of Lax matrices
	\begin{equation}
	\mathbf{L}_{s}(\lambda)=\ii 
	\begin{pmatrix}
	[\ii \lambda+\gamma\ \mathbf{s}^{z}]_q &  \mathbf{s}^{-} \cr
	\mathbf{s}^{+} & [\ii \lambda- \gamma\ \mathbf{s}^{z}]_q,
	\end{pmatrix}
	\label{eqn:Lax_operator_deformed}
	\end{equation}
	where tensor product is taken over physical space and matrix multiplication and trace over auxiliary space
	\begin{equation}
	T_s(\lambda)=\tr{(\mathbf{L}_s(\lambda)^{\otimes n})}.
	\end{equation}
	The generator $ \mathbf{s} $ satisfy $ \mathcal{U}_q(\mathfrak{sl}_2) $ relations and unitary representations read
	\begin{align}
	\mathbf{s}^{\z}\ket{n} &= (s-n)\ket{n},\\
	\mathbf{s}^{+}\ket{n} &= \sqrt{[2s-n]_{q}[n+1]_{q}}\ket{n+1},\\
	\mathbf{s}^{-}\ket{n+1} &= \sqrt{[2s-n]_{q}[n+1]_{q}}\ket{n},
	\label{eqn:compact_representation}
	\end{align}
	with $ q-$deformation defined as $ [\bullet]_q=\frac{\sinh(\bullet \gamma)}{\sinh \gamma} $, representation parameter $ s $ being half-integer and $
	n\in\{0,2 s\} $. To evaluate kernels \eqref{ker1} we introduce double Lax matrices
	\begin{equation}
	\mathbb{L}^{\pm}_{s}(\lambda,\mu) =
	\mathcal{N}^{\pm}_{s}(\lambda,\mu)(\mathbf{L}^{\mp}_{s}(\lambda)\otimes \one_{s})(\one_{s} \otimes \mathbf{L}^{\pm}_{s}(\mu))
	=\sum_{\alpha \in \mathcal{J}}\mathbb{L}^{\pm \alpha}_{s}(\lambda,\mu)\sigma^{\alpha},
	\label{eqn:double_Lax}
	\end{equation}
	with a normalizing factor
	\begin{equation}
	\mathcal{N}^{\pm}_{s}(\lambda,\mu) = \left(L^{[\mp(2s+1)]}_{0}(\lambda) L^{[\pm(2s+1)]}_{0}(\mu)\right)^{-1},
	\label{eqn:normalization}
	\end{equation}
	and four spin transfer matrices
	\begin{equation}
	\mathbb{T}_{s,s'}(\lambda,\lambda^{\prime},\mu^{\prime},\mu) =
	(\mathbb{L}^{+}_{s}(\lambda,\lambda^{\prime})\otimes \one^{\otimes 2}_{s^{\prime}})
	(\one^{\otimes 2}_{s}\otimes \mathbb{L}^{+}_{s^{\prime}}(\mu^{\prime},\mu))=\sum_{\alpha \in \mathcal{J}}\mathbb{T}^{\alpha}_{s,s'}(\lambda,\lambda^{\prime},\mu^\prime,\mu)\sigma^{\alpha}.
	\label{4mat}
	\end{equation}
	Taking into account the property \eqref{pr2} kernel \eqref{ker1} can be written in terms of modified auxiliary transfer matrices as
	\begin{align}
		K_{s,s'}^x(\lambda,\mu) &= \lim_{n\to \infty}\frac{1}{n}\Big\{[\partial_{\lambda^{\prime}}\partial_{\mu^{\prime}}
		\Tr\,\mathbb{T}^{0,x}_{s,s^{\prime}}(\lambda,\lambda^{\prime},\mu^{\prime},\mu)^{n}]_{\lambda^{\prime}=\lambda,\mu^{\prime}=\mu}
		\nonumber \\
		&-\big[\partial_{\lambda'}\Tr\,\mathbb{L}^{+0,x}_{s}(\lambda,\lambda^{\prime})^{n}\big]_{\lambda'=\lambda}
		\big[\partial_{\mu'}\Tr\,\mathbb{L}^{-0,x}_{s'}(\mu,\mu^{\prime})^{n} \big]_{\mu'=\mu}\Big\}.
		\label{eqn:K-kernel_def}
	\end{align}	
	We introduce a compact notation, omitting all of the indices and parameters of auxiliary transfer matrices, and substituting a partial derivative with respect to $ \mu' $ with $ ' $ and the one with respect to $ \lambda' $ with $ \cdot $.
	The expression for kernels in a compact form reads
	\begin{equation}
	K_{s,s'}^x(\lambda,\mu)=\lim_{n\to\infty}\frac{1}{n}(\tr(\TT^n_{s,s'})-\tr(\LL_s^n)\tr(\LL_{s'}^n))'^\cdot.
	\end{equation}
	The contribution to kernel can be divided into four parts
	\begin{eqnarray}
	K^x_{s,s'}(\lambda,\mu)&=&\lim_{n\to\infty}\sum_{k=0}^{n/2-1}\tr(\TT_{s,s'}^{n-k-2}\TT_{s,s'}'\TT_{s,s'}^{k}\dot{\TT}_{s,s'})+\sum_{k=0}^{n/2-2}\tr(\TT_{s,s'}^{n-k-2}\dot{\TT}_{s,s'}\TT_{s,s'}^{k}\TT_{s,s'}')+\nonumber\\
	&&+\tr(\TT_{s,s'}^{n-1} \dot{\TT}'_{s,s'})-n \tr(\LL_s^{(n-1)}\dot{\LL}_s)\tr(\LL_{s'}^{(n-1)}\LL_{s'}').
	\end{eqnarray}
	The contributions from sub-leading left and right eigenvectors in traces are exponentially small in $ n $ and the kernel reads
	\begin{eqnarray}
	\nonumber
	K^x_{s,s'}(\lambda,\mu)&=&\lim_{n\to\infty}\sum_{k=0}^{n/2-1}\langle L_{s,s'}|\TT_{s,s'}'\TT_{s,s'}^{k}\dot{\TT}_{s,s'}|R_{s,s'}\rangle+\sum_{k=0}^{n/2-2}\langle L_{s,s'}|\dot{\TT}_{s,s'}\TT^{k}_{s,s'}\TT'_{s,s'}|R_{s,s'}\rangle+\\
	&&+\langle L_{s,s'}| \dot{\TT}'_{s,s'}|R_{s,s'}\rangle-n \langle l_s|\dot{\LL}_s|r_s\rangle \langle l_{s'}|\LL'_{s'}| r_{s'}\rangle+\mathcal{O}(\exp(-\gamma n)),
	\label{kernTT}
	\end{eqnarray}
	where $ \langle l_s | $, $ |r_s\rangle  $ are left and right leading eigenvectors of $ \mathbb{L}_s $ and $ \langle L_{s,s'}| $, $ |R_{s,s'}\rangle $ are leading eigenvectors of $ \TT_{s,s'} $. The leading eigenvalues of $ \mathbb{L}_s $ and $ \TT_{s,s'} $ are $ 1 $.
	Most of the calculations involved in kernel \eqref{kernTT} can be carried out in terms of two-spin auxiliary transfer matrices $ \mathbb{L} $ alone.
	We will first discuss the computation of the first term in equation \eqref{kernTT}. The left four-spin eigenvector is decomposable in terms of two-spin eigenvectors $ \langle L_{s,s'}|=\langle l_s|\otimes\langle l_{s'}| $. Applying the derivative of $ \TT $ to the left eigenvector does not effect one of the two spin subspaces and as a consequence of the property 
	$\langle l_s|\vec{\vmbb{L}}^+_s(\lambda) = 0$,
	only the contributions from modified identity components are left
	\begin{equation}
	\langle L_{s,s'}|\TT'_{s,s'}\TT_{s,s'}^{k}\dot{\TT}_{s,s'}|R_{s,s'}\rangle=\langle l_s |\LL'_s \LL^k_s \otimes \langle l_{s'}| \dot{\TT} |R_{s,s'}\rangle
	\label{kerncon}
	\end{equation}
	Now we split the above contribution into two parts
	\begin{equation}
	\langle l'|\equiv \langle l_{s'}|\LL_{s'}',\ \langle l'|=\langle l'_\perp|+(\langle l_s|\otimes\langle l'|)|R_0\rangle \langle l_{s'}|.
	\label{split}
	\end{equation}
	Inserting the second contribution of \eqref{split} into \eqref{kerncon} we get
	\begin{equation}
	(\tfrac{n}{2})\langle l_s|\otimes \langle l_{s'}|\LL'|R_{s,s'}\rangle \langle l_s|\dot{\LL}\otimes \langle l_{s'}\|R_{s,s'}\rangle=(\tfrac{n}{2})\langle l_s|\dot{\LL}|r_s\rangle \langle l_{s'}|\LL'| r_{s'}\rangle.
	\label{12prisp}
	\end{equation}
	The equivalence of r.h.s. and l.h.s. is conjectured and was checked for couple of instances.
	To calculate the contribution from the first term in eq. \eqref{12prisp} the geometric series has to be summed up 
	\begin{equation}
	\langle l'_\perp|\sum_{k=0}^\infty\LL_s^k=\langle l'_\perp|(\one-\LL_s)^{-1}=\langle \tilde{l'} |.
	\end{equation}
	$ \langle \tilde{l'}| $ can be obtained by solving the system of linear equations. The second term in \eqref{kernTT} can be reduced in complete analogy, yielding the following expression for kernel
	\begin{equation}
	K^x_{s,s'}(\lambda,\mu)=\langle\tilde{l'}|\otimes\langle l_{s'}|\dot{\TT}_{s,s'}|R_{s,s'}\rangle+\langle l_s|\otimes\langle\dot{\tilde{l}}|\TT'_{s,s'}|R_{s,s'}\rangle+\langle L_{s,s'}| \dot{\TT}'_{s,s'}|R_{s,s'}\rangle-\langle l_s|\dot{\LL}_s|r_s\rangle \langle l_{s'}|\LL'_{s'}| r_{s'}\rangle.
	\label{kernTT2}
	\end{equation}
	To simplify calculations of the kernel \eqref{kernTT2} $ U(1) $ symmetry  can be employed.
	
	Using the above prescription one can obtain an exact expression for kernel of local charges in isotropic point
	\begin{equation}
	K^x_{1/2,1/2}(\lambda,\mu)=(1-x^2)\frac{(3+x^2(3+2(\lambda-\mu)^2+2\lambda \mu))}{4(1+\lambda^2)(1+\mu^2)(1+(\lambda-\mu)^2)}
	\label{kernel1}
	\end{equation}
	We conjecture, following the numerical results, that exact form of kernel for $ x=0 $ and any anisotropy $ \Delta $ takes the following form
	\begin{equation}
	K_{1/2,1/2}^0(\mu,\lambda)=\frac{\sinh ^4(\gamma ) (\cosh (2 \gamma )+\cos (2 (\lambda -\mu ))+\cos (2 \lambda )+\cos
		(2 \mu )+2)}{(\cos (2 \lambda )-\cosh (2 \gamma )) (\cosh (2 \gamma )-\cos (2 \mu ))
		(\cos (2 (\lambda -\mu ))-\cosh (2 \gamma ))}.
	\end{equation}
	
	Similar derivation leads to the expression for overlaps of current with charges
	\begin{equation}
	J^x_s(\lambda)=\lim_{n\to\infty}\frac{1}{n}\langle J,X_s(\lambda)\rangle ^x=\langle j_1,X_s(\lambda)\rangle^x.
	\end{equation}
	The connected part of the inner product is $ 0 $ while the expression consists of two contributions up to exponential corrections from sub-leading eigenvalues
	\begin{equation}
	J^x_s(\lambda)= \frac{\ii}{4} (1-x^2) (\langle l_s|(\LL_s^{++}\LL_s^{+-}-\LL_s^{+-}\LL_s^{++})'|r_s\rangle+\langle \tilde{l}_s'|(\LL_s^{++}\LL_s^{+-}-\LL_s^{+-}\LL_s^{++})|r_s\rangle).
	\end{equation}
	For isotropic point we obtained exact expressions for overlaps up to $ s=5 $. Here we list few of the lowest ones
	\begin{eqnarray}
	J^x_{1/2}(\lambda)&=&\frac{\lambda x \left(1-x^2\right)}{2\left(\lambda ^2+1\right)^2}\\
	J^x_{1}(\lambda)&=&\frac{384 \lambda  x \left(1-x^2\right)}{2\left(4 \lambda ^2+9\right)^2 \left(x^2+3\right)^2}\\
	J^x_{3/2}(\lambda)&=&\frac{\lambda  x \left(1-x^2\right) \left(x^4+2 x^2+5\right)}{2\left(\lambda ^2+4\right)^2
		\left(x^2+1\right)^2}\\
	J^x_{2}(\lambda)&=&\frac{640 \lambda x \left(1-x^2\right) \left(5 x^4+6 x^2+5\right)}{2\left(4 \lambda
		^2+25\right)^2 \left(x^4+10 x^2+5\right)^2}
	\end{eqnarray}
	Notice that in all of the above expression the overlap takes the separable form
	\begin{equation}
	J^x_s(\lambda)=\frac{\lambda}{2((s+\half)^2+\lambda^2)^2}g_s(x).
	\end{equation}
	We conjecture that in general the leading term of  overlaps in $ x $ reads
	\begin{equation}
	J^x_{s}(\lambda)=\frac{(2s+1)^2-1}{6}\frac{x \lambda}{((s+\half)^2+\lambda^2)^2}+\mathcal{O}(x^2).
	\label{olap1}
	\end{equation}
	In the $ XXZ $ case the overlaps take a similar form. More precisely they seem to be decomposable into three contributions
	\begin{equation}
	J^x_s(\lambda)=\frac{\sin (2 \lambda)\sinh^2\gamma}{2(\cosh((2 s+1)\gamma)-\cos(2 \lambda))^2}g_s(x)\tilde{g}_s(\gamma	),
	\label{olapform}
	\end{equation}
	Calculating few of the lowest overlaps leads us to conjecture that the function $ \tilde{g} $ takes the following form
	\begin{equation}
	\tilde{g}_{s}(\gamma	)=\frac{\sinh((1+2 s)\gamma)}{\sinh\gamma}.
	\end{equation}
	Note that the above conjecture offers an expression for an overlap at any $ \Delta $ provided that the expression for $ \Delta=1 $ is known. Thus it also yields a conjecture for overlaps in the vicinity of $ m=0 $
	\begin{eqnarray}
	J_s^x(\lambda)=\frac{1}{3}((2s+1)-(2s+1)^{-1})\frac{\sinh((2s+1)\gamma)}{\sinh\gamma}\frac{\sin(2\lambda)x\sin^2\gamma}{\cosh(2 s+1)-\cos(2 \lambda)}+\mathcal{O}(x^2).
	\end{eqnarray}
	\section{E: Derivation of the lower bound on diffusion in Heisenberg model from all local charges}
	In the anisotropic case we can reduce the calculation to a finite integration interval to $ [-\frac{\pi}{2},\frac{\pi}{2}] $ since the kernels and overlaps are periodic with a period $ \pi $.
	We are interested in behavior of Drude weight only in the vicinity of half-filling $ x=0 $. If we assume analyticity of kernels and overlaps the equation \eqref{sys} from the main text in leading order reads
	\begin{equation}
	\sum_{s'=1/2}^{\infty}\int K^0_{s,s'}(\lambda,\mu) h'_{s'} (\mu) d\mu=J_s'(\lambda).
	\label{eqts}
	\end{equation}
	where $J'_s(\lambda):= \partial_x J_s^x(\lambda)|_{x=0} $ and $ K^0_{s,s'}(\lambda,\mu) $ are Hilbert-Schmidt kernels at half-filling. The solution of equation \eqref{eqts} provides a lower bound on Drude weight in the vicinity of half-filling
	\begin{equation}
	\langle \bar{J}, j\rangle^{x}\geq x^2\sum_{s,s'=\half}^\infty\int d\lambda \int d\mu\  K^0_{s,s'}(\lambda,\mu)h'_s(\lambda)\bar{h}'_{s'}(\mu)+\mathcal{O}(x^4),
	\label{HSKmm1}
	\end{equation}
	where $ \bar{J} $ is time average of current.
	Taking into account only local conserved charges the Fredholm equation in first order in $ x $ reads
	\begin{eqnarray}
	&&\int_{-\pi/2}^{\pi/2}\frac{\sinh ^4(\gamma ) (\cosh (2 \gamma )+\cos (2 (\lambda -\mu ))+\cos (2 \lambda )+\cos
		(2 \mu )+2)}{(\cos (2 \lambda )-\cosh (2 \gamma )) (\cosh (2 \gamma )-\cos (2 \mu ))
		(\cos (2 (\lambda -\mu ))-\cosh (2 \gamma ))} h'_{1/2}(\mu)d\mu \nonumber\\		
		&&\qquad\qquad\qquad\qquad\qquad\qquad=\frac{\sinh(2\gamma)}{\sinh\gamma}\frac{\sin(2\lambda) \sin^2\gamma}{2(\cosh (2\gamma)-\cos(2 \lambda))^2}.  \label{fhgamma}
	\end{eqnarray}
	Introducing a function  
	\begin{equation}
	\tilde{h}_{1/2}(\mu)=\frac{\sinh^3\gamma\ h'_{1/2}(\mu)}{ \sinh(2\gamma) (\cosh(2\gamma)-\cos(2\mu))},
	\end{equation}
	and rescaling the arguments $ 2\mu\to\mu  $, $ 2\lambda\to\lambda $ equation \eqref{fhgamma} reduces to
	\begin{equation}
	\int_{-\pi}^{\pi}\frac{\cosh(2\gamma)+\cos(\lambda-\mu)+\cos\lambda+\cos\mu+2}{\cosh(2\gamma)-\cos(\lambda-\mu)}\ \tilde{h}_{1/2}(\mu)d\mu=\frac{\sin\lambda}{\cosh (2\gamma)-\cos( \lambda)}.
	\label{lsys}
	\end{equation}
	To solve the above equation we expand the function $ \tilde{h} $ in terms of Fourier modes $ \tilde{h}_{1/2}(\mu)=\sum_{j=1}^{\infty}\alpha_j\sin(j\mu) $.
	The integration of the l.h.s. of the above expression can be carried out after observing the identity
	\begin{equation}
	\int_{-\pi}^{\pi}\frac{\cos (k \mu)}{\cosh(2\gamma)-\cos \mu} \dd \mu=2 \pi \frac{e^{-2\gamma k}}{\sinh( 2 \gamma)}.
	\label{ident}
	\end{equation}	
	A set of equations arising from equation \eqref{ident} corresponding to distinct Fourier modes reads
	\begin{equation}
	\alpha_k+\alpha_{k-1}\frac{e^{2\gamma}}{2(e^{2\gamma}+1)}+\alpha_{k+1}\frac{1}{2(e^{2\gamma}+1)}=\frac{e^{2\gamma}-1}{2\pi(e^{2\gamma}+1)},\ \ \alpha_0=0.
	\label{sys1}
	\end{equation}
	To solve the above system we introduce new variables $ \alpha_k=\alpha_\infty+\tilde{\alpha}_k $, where $ \alpha_\infty $ is determined by the condition
	\begin{equation}
	\alpha_\infty+\alpha_{\infty}\frac{e^{2\gamma}}{2(e^{2\gamma}+1)}+\alpha_{\infty}\frac{1}{2(e^{2\gamma}+1)}=\frac{e^{2\gamma}-1}{2\pi(e^{2\gamma}+1)}\to\alpha_{\infty}=\frac{e^{2\gamma}-1}{3\pi (e^{2\gamma}+1)}.
	\end{equation}
	Plugging the ansatz for $ \tilde{\alpha} $ into eq. \eqref{sys1} yields a modified system of linear equations
	\begin{equation}
	\tilde{\alpha}_k+\tilde{\alpha}_{k-1}\frac{e^{2\gamma}}{2(e^{2\gamma}+1)}+\tilde{\alpha}_{k+1}\frac{1}{2(e^{2\gamma}+1)}=\frac{e^{2 \gamma } \left(e^{2 \gamma }-1\right)}{6 \pi  \left(e^{2 \gamma }+1\right)^2}\delta_{k,1}.
	\label{sys2}
	\end{equation}
	Using an ansatz $\tilde{\alpha}_k= \delta b^k$ we obtain the following solution 
	\begin{equation}
	\tilde{\alpha}_k=\frac{\left(1-e^{2 \gamma }\right) ( \sqrt{1+e^{2\gamma}+e^{4\gamma}}-1-e^{2\gamma})^{k}}{3 \pi  \left(e^{2 \gamma }+1\right)}.
	\end{equation}
	The resulting expression can be inserted in equation \eqref{HSKmm1}, yielding 
	\begin{equation}
	D(x)\geq\frac{x^2 \sinh^2(2\gamma)}{4\sinh^2\gamma}\int_{-\pi}^{\pi}\frac{\sin\mu}{\cosh 2 \gamma-\cos\mu}\tilde{h}_{1/2}(\mu)d\mu+\mathcal{O}(x^4).
	\end{equation}
	Finally, after using the relation \eqref{ident} we obtain the following lower bound
	\begin{equation}
	D(x)\geq \frac{x^2}{3}\cosh(\gamma)\left(e^{-\gamma}+\frac{2 \sinh \gamma}{\sqrt{1+e^{2\gamma}+e^{4\gamma}}+2+e^{2\gamma}}\right)+\mathcal{O}(x^4).
	\end{equation}
	Note that in the process of deriving the above expression non-convergent series expansion of $ h_{1/2}(\lambda) $ has been integrated term by term yielding finite result. The derivation can be made rigorous by introducing a finite set of charges $ \tilde{X}_k=\int_{-\pi/2}^{\pi/2} d\lambda\ \sin (2 k \lambda)(\cosh(2 \gamma)-\cos(2 \lambda))) X_{1/2}(\lambda),\ k\in\{1,2,...,n\} $, finding an optimal lower bound for a given $ n $ and taking limit $ n\to\infty $ at the end.
	\section{F: Lieb-Robinson velocity}
	Here we provide some numerical results regarding the spreading of spatio-temporal spin current-current correlation functions in Heisenberg model for $ \Delta=1.5 $. The velocity with which correlations spread is upper bounded by Lieb-Robinson velocity. In Fig.~\ref{fig2} we plot the current-current spatio-temporal correlation functions and the dependence of integrated cone restricted correlation function
	\begin{equation}
	C(v)=\int_{0}^{T}dt \sum_{x=-\lceil 1+ v t \rceil}^{\lceil 1+ v t \rceil} \langle \tau_t(j) j_x\rangle
	\label{odv}
	\end{equation}
	on velocity $ v $. For the estimate of velocity at which the correlations spread we take $ v $ at which $ C(v) $ becomes almost constant.
	\begin{figure}
		
		\includegraphics[width=\linewidth]{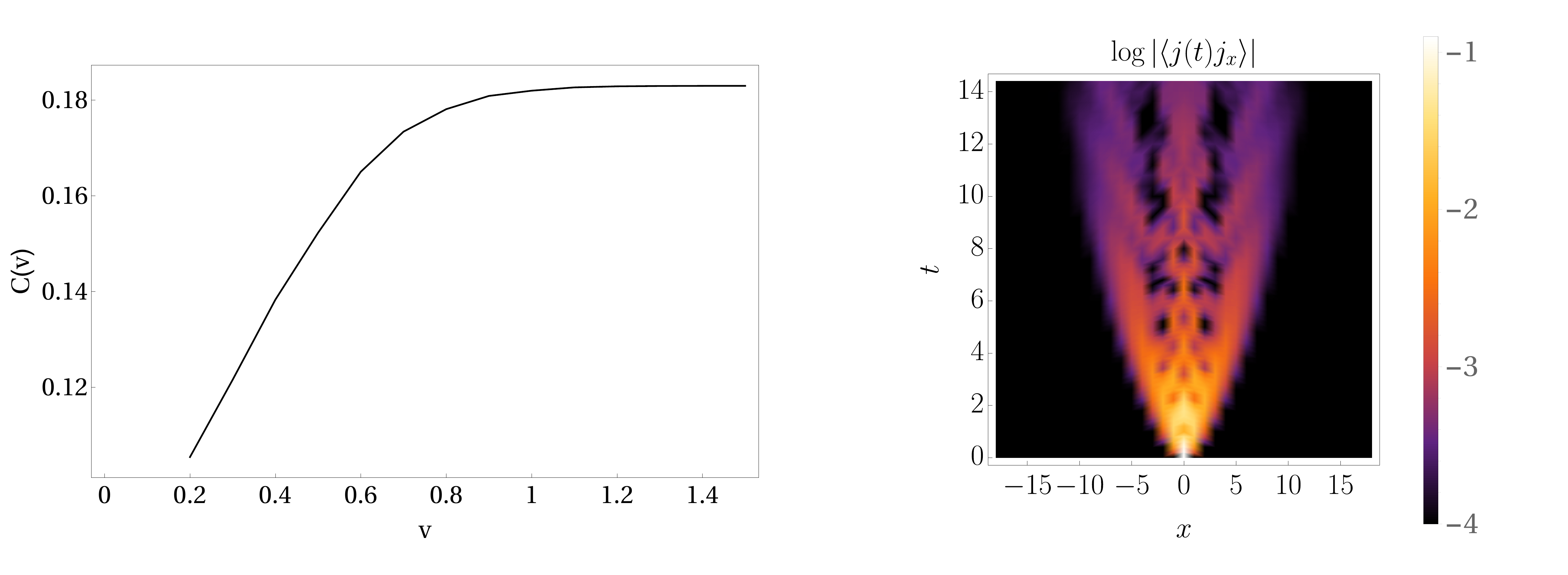}
		\center
		\caption{
			\label{fig2} 
			On the left: the dependence of spatio-temporally integrated current-current correlation function on the velocity \eqref{odv}. On the right: spatio-temporal current-current correlation function in log scale.}
		
	\end{figure}
	From Fig.~\ref{fig2} we can conclude that the velocity of the spread of correlations is $v \approx 4 $. This velocity is consistent with the rate of propagation of rays in spatio-temporal autocorrelation functions which can be seen in Fig.~\ref{fig2}.
	\section{G: List of assumptions}
	Our derivation of the lower bound on diffusion constant could be claimed rigorous provided we make certain assumptions. We believe there is no doubt that all these assumptions are justified in typical 
	physical models and in certain classes of generic models they can be simply stated as facts (or independently proven). Here we spell out the complete list of assumptions that have been made:
	\begin{enumerate}
		\item Finiteness of thermal correlation length and quasilocality (i.e. boundedness in the operator norm) of the imaginary time propagation of the current $ \tau_{i\lambda}(j) $, for any $\lambda \in [0,\beta]$, or in precise terms
		$$
		\lim_{r\to\infty}\lim_{n\to\infty}\int_{-T^r}^{T^r} \dd t \, \langle \tau^r_t(j), J_r\rangle_n^\beta = \lim_{r\to\infty}\int_{-T^r}^{T^r} \dd t \, \langle \tau^r_t(j), J_r\rangle_r^\beta,
		$$
		where $ T^r=\frac{r}{\alpha v_{\rm LR}} $, $ \alpha>1 $, and $v_{\rm LR}$ is the Lieb-Robinson velocity.
		\item Analyticity at $x=0$ of $ D(\beta,x)$
		\item Analyticity at $x=0$ of finite $n$ corrections to $ D(\beta,x)$ up to (including) $ \frac{1}{n} $ terms,
		\item Analyticity at $x=0$ of free energy $ \beta F_n(x,\beta)=-\log \tr (P_n^{(x+1)n} e^{-\beta H_n})$
		\item Disregarding the states which are more than $x^*=(2 n)^{1/2+\varepsilon}$ away from half-filling for $0 < \varepsilon < 1/4$. The relative fraction of these states is of the order $ \mathcal{O}(\exp(-A n^{2\varepsilon})) $.
	\end{enumerate}

	
	\end{widetext}
\end{document}